# Magnetic behavior of $Gd_3Ru_4Al_{12}$, a layered compound with distorted Kagome net

Venkatesh Chandragiri, Kartik K Iyer and E.V. Sampathkumaran
*Tata Institute of Fundamental Research, Homi Bhabha Road, Colaba, Mumbai 400005, India*

Abstract

The magnetic behavior of the compound, $Gd_3Ru_4Al_{12}$, which has been reported to crystallize in a hexagonal structure (space group *P6$_3$/mmc*) about two decades ago, had not been investigated in the past literature despite interesting structural features (that is, magnetic layers and triangular as well as Kagome-lattice features favoring frustrated magnetism) characterizing this compound. We report here the results of magnetization, heat-capacity and magnetoresistance studies in the temperature ($T$) range 1.8 – 300 K. The results establish that there is a long-range magnetic order of an antiferromagnetic type below ($T_N=$) 18.5 K, despite a much larger value (~ 80 K) of paramagnetic Curie temperature with a positive sign characteristic of ferromagnetic interaction. We attribute this to geometric frustration. The most interesting finding is that there is an additional magnetic anomaly below ~55 K before the onset of long range order in the magnetic susceptibility data. Concurrent with this observation, the sign of isothermal entropy change, *ΔS= S(0)-S(H)*, where *H* is the externally applied magnetic field, remains positive above $T_N$ with a broad peak above $T_N$. This observation indicates the presence of ferromagnetic clusters before the onset of long range magnetic order. Thus, this compound may serve as an example for a situation in which magnetic frustration due to geometrical reasons is faced by competition with such magnetic precursor effects. There is also a reversal of the sign of *-ΔS* in the curves for lower final fields (*H*<30 kOe) on entering into magnetically ordered state consistent with the entrance to antiferromagetic state. The magnetoresistance behavior is consistent with above conclusions.

*JPCM (in press)*

I. **INTRODUCTION**

In solid state physics, materials containing layers of magnetic ions or those with topologically frustrated magnetism have been attracting a lot attention for the past few decades. Among the latter, the topic of investigations of those with antiferromagnetically coupled magnetic ions in a triangular arrangement and in Kagome lattices has been great interest. The rare-earth (R) family of the type $R_3Ru_4Al_{12}$ is an interesting one among intermetallics, as the structure contains layers, as well as triangular and Kagome lattice features [1, 2]. Despite such fascinating crystallographic features, the reports on the magnetic behavior of the members of this rare-earth family are sporadic over a period of nearly two decades. It appears that the interest in this family is rekindled in recent years with the reports on R= Tb and Dy members proposing antiferromagnetic ordering below ($T_N$=) 22 K and 7 K respectively [3, 4], and ferromagnetism in Pr and Nd members at relatively higher temperatures (below 40 K). While Ce in the Ce compound is found to remain intermediate valent [5, 6], Yb analogue was found to be antiferromagnetic below 1.5 K [7]. Investigations based on single crystals of Tb and Dy compounds revealed jumps in isothermal magnetization data and the magnetic properties were found to be strongly anisotropic, with a non-collinear magnetic structure with the R moments (in the case of Dy) [3, 4]. Incidentally, the U analogue was also subjected to some magnetic investigations and this compound is characterized by a $T_N$ value of 9.5 K with a localized behavior of 5f electrons and a confinement of the magnetic moments to the basal plane with a non-collinear structure [8, 9]. Surprisingly, to our knowledge, the Gd compound was not paid any attention for its magnetic behavior, though this turned out to be the prototype for this family of f-electron systems [1]. We therefore carried out magnetization (*M*), heat-capacity (*C*) and electrical resistivity (*ρ*) measurements, the results of which are reported in this article.

II. **Crystallographic details**

The compound under discussion crystallizes in a hexagonal structure in the space group *P6₃/mmc*, in which there are two kinds of layers: planar layers of $Gd_3Al_4$ and puckered layers of $Ru_4Al_8$ stacked alternately along the *c*-axis (see figure 1a). Different views of atomic arrangements are shown in figure 1. Crystallographic features at great length have been described in Refs. 1 and 2 and we present here a brief description of essential features. There is one site (*6(h)*) for Gd, two sites for Ru (*2(a), 6(g)*), and four sites for Al (*12(k): Al1; 6(h): Al2; 4(f): Al3; 2(b): Al4*). Gd atoms form distorted Kagome nets with larger and smaller triangles with these triangles sharing vertices and with different edge lengths (see figure 1b). These structural features naturally favour geometrically frustrated magnetism in the event that intersite interaction is antiferromagnetic. Some Al2 atoms are arranged in equilateral triangles inside the hexagons of Kagome nets. Al4 occupy the center of the large Gd-triangles (see figure 1b). In the adjacent layer (see figure 1a), Ru atoms form a triangular mesh centered by Al (figure 1c), with Al3 located approximately in the plane and Al1 above and below this Ru layer (figure 1d), thereby explaining puckered nature of the $Ru_4Al_8$ layers; but Ru atoms are not adjacent to each other within the layer, but have only Al as neighbours. These puckered $Ru_4Al_8$ layers are at *z*= 0 and 0.5 and Gd layers are at *z*= 0.25 and 0.75.

III. **EXPERIMENTAL DETAILS**

The polycrystals of the compound, $Gd_3Ru_4Al_{12}$, was prepared by arc melting several times stoichiometric amounts of high purity (>99.9%) constituent elements in an atmosphere of argon. The ingot was subsequently annealed at 800 C in an evacuated quartz tube for a week. X-ray diffraction (XRD) pattern confirms that the sample forms in a single phase, which was further verified by scanning electron microscopy. Rietveld fitted XRD pattern is shown in figure 2 and the parameters obtained are shown in the plot. The lattice constants obtained with this fitting compare quite well with those reported in the literature [2]. Temperature (*T*) dependence of magnetic susceptibility (χ) was measured (1.8 – 300 K) with the help of a commercial (Quantum Design) SQUID magnetometer for zero-field-cooled (ZFC) and field-cooled (FC) conditions of the specimen in the presence of a few selected magnetic fields (*H*); *M(H)* curves were also obtained at several temperatures for ZFC condition (cooled from 300 K). *ρ(T)* curves in the presence of several magnetic fields and *ρ(H)* curves at several temperatures were obtained with the help of a commercial Physical Property Measurement System, PPMS (Quantum Design); heat-capacity behavior was also studied with the same PPMS. While presenting the results in the form of figures, the experimental data points in some cases are omitted for the sake of clarity and only a smooth line through the data points are shown.

## IV. RESULTS AND DISCUSSION
### A. Magnetic susceptibility behavior

Figure 3a shows $\chi$ as a function of $T$ measured in a field of 5 kOe. The plot of $\chi^{-1}$ suggests that Curie-Weiss behavior is obeyed in a narrow temperature range, that is, above about 200 K. This deviation from Curie-Weiss behavior below 200 K is attributed to gradual dominance of short-range magnetic correlations with decreasing temperature (before the onset of antiferromagnetism, *see below*). The effective moment obtained from the Curie-Weiss region turns out to be ~8.1 $\mu_B$/Gd, which is very close to the free ion value of $Gd^{3+}$ (7.94 $\mu_B$), thereby establishing that possible magnetic moment on Ru, if present, could be very small. The value of paramagnetic Curie temperature ($\theta_p$) is found to be about +80 K with the positive sign establishing that the interaction between Gd ions is actually ferromagnetic. However, following a monotonic increase of $\chi$ with decreasing temperature till 20 K, there is a distinct peak at 18.5 K (see the $\chi$ curve for $H$= 5 kOe in figure 3a), which indicates onset of antiferromagnetic ordering. This observation and the reduced value of the magnetic ordering temperature ($T_N$= 18.5 K) with respect to the value of $\theta_p$ imply magnetic frustration. We take this as a signature of a competition with antiferromagnetic interaction, which leads to geometric frustration. We find that the $\chi(T)$ curves obtained for ZFC and FC conditions at low fields (say, 100 Oe) nearly coincide, thereby ruling out spin-glass freezing in this compound, as shown in the mainframe of figure 3b. What is interesting is that, unlike the behavior of the curve obtained in 5 kOe, there is a sharp increase of $\chi$ near 55 K in the curve obtained in 100 Oe, followed by a relatively more gradual increase till $T_N$. Thus, there is an additional subtle magnetic anomaly before the onset of long range magnetic order. Incidentally, we have also performed isothermal remnant magnetization, memory experiments and ac susceptibility measurements and we did not find the characteristics of spin-glasses in any of these properties below 60 K; the plot *ac* susceptibility versus temperature is similar to that seen in *dc M* measurements with 100 Oe.

In order to understand the 55 K feature, the behavior of $\chi$ versus $T$ is shown in figures 3b (inset) and 3c for several fields. It is clear that the upturn near 55 K in $\chi$ gets smeared with increasing $H$. This behavior mimics that expected for Griffiths phase, an idea which was proposed several decades ago [11] to describe a situation in which long range ferromagnetic order is suppressed to a temperature far below $\theta_p$ (due to some reasons like disorder), resulting in an intermediate $T$-region in which ferromagnetic clusters are dispersed in a paramagnetic matrix; in this phase, there is no global spontaneous magnetization. Since these ferromagnetic clusters are characterized by larger effective spins than individual magnetic ions, inverse $\chi$ shows (see figure 3c) dramatic deviation from the high temperature Curie-Weiss behavior at the formation of Griffiths phase. In the Griffiths phase,

$$\chi^{-1}(T) \text{ is proportional to } (T-T_G)^{1-\lambda}$$

where $\lambda$ takes a value between 0 and 1 and $T_G$ is called Griffiths temperature [12, 13]. The $\lambda$ naturally takes a value of 1 above $T_G$. In order to demonstrate this behavior, we have plotted $\chi^{-1}(T)$ for several values of $H$ in log-log scale in figure 3c. The fitting for the data taken with 100 Oe above and just below $T_G$ is shown in the inset. The value of $\lambda$ is found to be close to 1 above $T_G$. The value as the temperature is lowered turns out to be about 0.8, close to that expected for Griffiths phase. In short, as the temperature is lowered, Griffiths-phase-like feature seems to appear before the onset of long range antiferromagnetic order in this compound. (For a few other examples for Griffiths phase behavior, see Refs. 14-16, though direct experimental proof is not easy to obtain).

A word of caution is required for the above hypothesis. Naively speaking, Griffiths phase is expected in randomly diluted systems, and, it is therefore not clear how such a phase is possible in concentrated magnetic systems. If this is the explanation for our observation, possibly crystallographic defects and/or geometrically-induced magnetic frustration may also trigger the formation of such a phase. Confirmation of this hypothesis therefore requires further research. In any case, the present results render a strong evidence for a recent proposal [17] that new collective mechanisms develop in order to fight geometrically frustrated magnetism before long range ordering sets in at a lower temperature. In this connection, we would like to add that, even in the case of Dy analogue, which was proposed to undergo Néel order below 7 K [4], we find a magnetic feature at low fields above $T_N$ around 18 K which is nearly the same as $\theta_p$ [18]. The fact that the temperature at which this new feature appears is intimately linked to $\theta_p$, moving with $T_N$, suggests that it is essentially related to rare-earth magnetism (and hence at present we tend to rule out the role of possible Ru magnetism).

### B. Isothermal magnetization and isothermal entropy change behavior

We have also inspected the behavior of magnetocaloric effect as indicated by isothermal entropy change, -$\Delta S= S(0)-S(H)$, as the sign of this property is known to be sensitive to the magnetic state of the material [19]. For this purpose, we have measured isothermal $M$ at close intervals of temperature (typically 2 to 3 K) and representative curves are shown in figure 4a. It is distinctly clear that at very low temperatures (below 18 K), there is an upward curvature in the range 6-12 kOe (followed by a tendency for saturation at high fields), as though there is a meta-magnetic transition. There are also additional magnetic-field transitions near 20 and 35 kOe, as revealed by derivative plots shown (by vertical arrows) in the inset of figure 4a for 2K. The $M(H)$ plots are not hysteretic and there is no evidence for spontaneous magnetization. In figure 4b, we show the Arrott plot (the plots of $M^2$ versus $H/M$) and the curves are quite complex. The plots below $T_N$ exhibit a S-shaped behavior (that is, with a negative slope, see also the inset of figure 4b), which is a characteristic feature of magnetic-field-induced ferromagnetism [20]. Above $T_N$, negative slope is absent in the plots and the curves intercept on the x-axis (and there is no evidence for spontaneous magnetization). All these findings are consistent with antiferromagnetic ordering below 18.5 K. The magnetic moment at 50 kOe at 2K is close to 22 $\mu_B$/formula unit, which is marginally higher than that expected for 3Gd ions (21 $\mu_B$), which can be attributed to a small contribution from Ru conduction band.

Isothermal entropy change, $\Delta S [= S(H_2)-S(H_1)]$, for a change of the magnetic field from $H_1$ to $H_2$ was obtained from

$$\Delta S = \int_{H_1}^{H_2} \left(\frac{\partial M}{\partial T}\right)_H dH$$

on the basis of Maxwell thermodynamic relationship

$$\left(\frac{\partial S}{\partial H}\right)_T = \left(\frac{\partial M}{\partial T}\right)_H$$

The values of -$\Delta S$ thus derived are plotted in figure 4c below 100 K for selected values of final fields (5, 15, 30, and 50 kOe) with initial field being zero. It is clear that –$\Delta S$ values increase gradually with decreasing $T$ attaining a peak above $T_N$. The fact that these values are reasonably large (~5 J/kg-K at the peak for $H_2$= 50 kOe) spreading over a wide $T$-range is supportive of the existence of Gd ferromagnetic clusters. [Such a large magnitude of $\Delta S$ at the peak is not expected from possible weak magnetism from Ru]. At $T_N$, there is a sign crossover of -$\Delta S$, with the negative sign typical of antiferromagnetism. The temperature at which the sign crossover occurs gradually decreases with increasing $H$, indicating a suppression of $T_N$ (see also section IV.C) with $H$. At high fields, the values remain in the positive zone consistent with field-induced ferromagnetism. Incidentally, there is another upturn below 6 K; this presumably indicates the existence another magnetic transition well below 10 K and it is at present not clear whether Ru tends to undergo a weak magnetic ordering. This interpretation is strengthened by a similar proposal even for La analogue [5].

### C. Heat-capacity behavior

Figure 5a shows heat-capacity as a function of temperature. The zero-field data shows a lamda-type anomaly in the vicinity of $T_N$. With increasing temperature, after crossing $T_N$, there is a monotonic increase of $C$ (measured till 100 K). This is consistent with the fact that the ferromagnetic cluster formation proposed above is not a well-defined global phase transition. In order to understand this better, we have obtained magnetic part ($C_{mag}$) to $C$ from the knowledge of the heat-capacity values of the Y analogue [4]. The non-magnetic contribution (essentially phononic) was derived by multiplying $C(T)$ values of the Y compound by a factor of 0.903, which was obtained following the procedure of Ref. 4 and the corresponding Debye temperature ($\theta_D$) is estimated to be ~432 K. The plots of the non-magnetic part and the magnetic part derived are also shown in the figure 5a. It is distinctly clear that there is a broad peak in the vicinity of 55 K attributable to the magnetic phase above $T_N$. Also, the peak value near $T_N$ (~ 6 J/Gd mol K) is less than the value (20.15 J/Gd mol K) expected [21, 22] for any equal-moment magnetic structures (simple antiferro, ferro or helimagnetic). This implies that either full entropy is not released at $T_N$ or there is an amplitude modulation of the magnetic structure. To understand this, the magnetic entropy ($S_{mag}$) as a function of $T$ derived from the values of $C_{mag}$ are shown in the inset of figure 5a. We find that, at $T_N$, the value of magnetic entropy released (~20.2 J/mol K) is far less than that theoretically expected (51.87 J/mol K) for 3 Gd ions, and that full entropy is released at ~51 K, which is close to the temperature where the broad magnetic anomaly above $T_N$ prominently appears. We have obtained $C(T)$ in the presence external magnetic fields below 30 K, the results of which are shown in figure 5b in the form of $C/T$ versus $T$. The peak shifts to a lower temperature with the application of a magnetic field (even for 5 kOe) and this is consistent with antiferromagnetism. An application of higher magnetic fields (>10 kOe), which forces ferromagnetic alignment, smears out this peak. We have also derived isothermal entropy change, $\Delta S$, on the basis of $C(T)$ obtained in various fields, and the curves obtained (and also plotted in figure 4c) are qualitatively in good agreement with those obtained from magnetization data employing Maxwell relation, presented above. Before concluding section, we would like to add that there is a weak and broad curvature below 6 K in the plot of $C$ versus $T$ (figure 5b). Though it can be attributed to a Schottky-like anomaly in the magnetically ordered state involving quantum levels of (2J+1)-fold degenerate multiplet due to exchange field [22], we are not able to rule out interference from possible weak magnetism from Ru.

### D. Electrical resistance and magnetoresistance behavior

Temperature dependence of electrical resistivity is plotted in figure 6a. If one looks at the curve obtained in the absence of magnetic field, it is clear that the material exhibits metallic behavior with ρ decreasing gradually, with a very gradual change of slope below 200 K as in the case of Dy compound [4]. At the onset of long range magnetic order, there is a sudden fall due to the loss of spin-disorder contribution. We have fitted the data above $T_N$ to the Block-Grüneisen function as described in Ref. 4 for Dy case. We could get a $\theta_D$ value of about 400 K, only if the fit is restricted to a temperature range far above $T_N$, that is, 200-300 K. However, if one extends the fitting down to 60 K, we get a lower value of 285 K, which is in disagreement with that obtained as described in section IV.C. Clearly, short-range magnetic correlations, which tend to dominate below ~200 K gradually, as evidenced by inverse χ, distort the values of $\theta_D$. The data obtained in the presence of external magnetic fields are revealing. While the curves overlap in the higher temperature region (above ~100 K), there is a visible deviation from the zero-field curves at lower temperatures with ρ attaining lower resistivity values. The feature around $T_N$ is also smoothened as that observed in many materials at the magnetic transition temperature. It is to be noted that the sign of magnetoresistance, MR [defined as $\{\rho(H)-\rho(0)\}/\rho(0)$], is negative (see Fig. 6b), but the temperature near which the magnitude is clearly visible is comparable to that of $\theta_p$, as though the short-range magnetic correlations setting in at such high temperatures could be of a ferromagnetic type. It is a well-known fact that antiferromagnetism (without magnetic Brillouin-zone gap) and metallic state are characterized by positive MR contribution, whereas ferromagnetic alignment results in negative contribution. The magnitude of MR also gets gradually larger below 100 K with decreasing temperature, peaking at $T_N$, whereas the usual contribution of paramagnetic part is expected to be weaker [23] even in Gd alloys (carrying large magnetic moment). With respect to the MR behavior below $T_N$, a shoulder near 6K is clearly discernable in the MR($T$) curves, which can be correlated to another magnetic feature at this temperature and it is possible that Ru magnetism also plays a role.

We have also obtained isothermal MR curves till 140 kOe at selected temperatures above and below $T_N$ (see figure 7 for the behavior at 2, 8, 15 and 35 K) and the curves are found to be non-hysteretic and symmetric with respect to the origin for positive and negative values of $H$ (*not shown*). As seen in figure 7, the magnitudes are rather large at high fields, consistent with figure 6. A notable observation is that the MR for $T<T_N$ is of positive sign at low fields as distinctly seen in the curves of 2 and 8 K. After attaining a peak, the MR tends to decline in its magnitude in the vicinity of 10 kOe, followed by a sharp change and a sign reversal at a higher field (near 20 kOe). As shown in figure 4, there is a magnetic-field induced magnetic transition near 10 kOe and therefore the observed MR behavior is a manifestation of this transition. We therefore attribute the observed shape of the MR curves to a competition between these contributions. For 35 K, which is well above $T_N$, well-known quadratic dependence with $H$ expected for paramagnetic state is not seen; instead, there is a complex shape of the MR($H$) curve with a negative sign in the entire field range of investigation and with a large magnitude at high fields. This may have to be attributed to ferromagnetic clusters.

## V. CONCLUSION

The results establish that the compound $Gd_3Ru_4Al_{12}$, with coexistence of magnetic layers, triangles and distorted Kagome nets, appears to be a geometrically frustrated magnetic system with Néel temperature of 18.5 K. There appears to be a competition between antiferromagnetism and ferromagnetism with varying $T$ and $H$. The most interesting finding is that the low-field magnetization exhibits an additional magnetic anomaly below 55 K before the onset of long-range magnetic order (mimicking the behavior of Griffiths phase). This finding reinforces the assertion [17] that various collective magnetic interactions can compete with lowering temperature in the event that geometrical frustration dominates, before the material undergoes long-range antiferromagnetic ordering. We hope that this work on this prototype compound of this f-electron family triggers further work in this ternary family in the areas of layered compounds, triangular lattices, Kagome lattices and frustrated magnetism.

Figure 1:
Crystal structure of $Gd_3Ru_4Al_{12}$ to reveal (a) unit cell, (b) planar and distorted Kagome net of $Gd_3Al_4$ layer viewed along c-axis, (c) puckered layer of $Ru_4Al_8$ viewed along c-axis and (d) puckered layer of $Ru_4Al_8$ viewed along a-axis.

Figure 2:
X-ray diffraction pattern (Cu $K_\alpha$) of $Gd_3Ru_4Al_{12}$. The diffraction peaks are indexed. The parameters obtained from Rietveld analysis are included.

Figure 3:
For $Gd_3Ru_4Al_{12}$, (a) inverse susceptibility obtained in a field of 5 kOe is plotted; the dashed line through the data points at high temperatures is a result of Curie-Weiss fitting; (b) χ data obtained in the presence of several fields (100 Oe, 500 Oe, 1 kOe, 2 kOe, 5 kOe, and 30 kOe) are plotted in the inset; for $H$= 100 Oe, the curves obtained for ZFC and FC conditions are plotted in the mainframe; and (c) inverse susceptibility as a function of temperature for several fields (100 Oe, 500 Oe, 1 kOe, 2 kOe, 5 kOe and 10 kOe) in the log-log form; inset shows fitting as described in the text for the 100 Oe curve. Unless otherwise stated, the lines through the data points serve as guides to the eyes. The arrows cutting across the curves in (b) and (c) correspond to the $H$ values mentioned above (in increasing order). For some curves, only a line through the data points are shown for the sake of clarity.

Figure 4:
Isothermal magnetization (in the unit of Bohr magneton per formula unit (f.u.)) behavior of $Gd_3Ru_4Al_{12}$ at selected temperatures and continuous lines are drawn through the data points to serve as guides. Derivative plot for 2 K is shown in the inset and vertical arrows are drawn to bring out magnetic-field induced transitions. (b) Arrott plot with the inset showing some region in the expanded form for selected temperatures to highlight the existence of negative slope for the sake of clarity. (c) Isothermal entropy change obtained from magnetization data as described in the text for three selected final values (5, 15, 30, and 50 kOe); the lines through the data points serve as guides to the eyes. The behavior obtained from heat-capacity data for the final fields 5, 30, and 50 kOe (with initial zero field) are also shown by continuous lines (omitting data points) for comparison. For some curves, only a line through the data points are shown for the sake of clarity.

Figure 5:
(a) Heat-capacity as a function of temperature for $Gd_3Ru_4Al_{12}$. The magnetic and non-magnetic part derived from the knowledge of heat-capacity of Y analogue (see text) are also shown by a dashed line. Inset shows magnetic entropy ($S_{mag}$). (b) Heat-capacity below 25 K obtained in the presence of several magnetic fields.

Figure 6:
(a) Electrical resistivity as a function of temperature for $Gd_3Ru_4Al_{12}$, obtained in the presence of several externally applied magnetic fields; (b) magnetoresistance, MR (={ρ(H)-ρ(0)}/ρ(0)), as a function of temperature.

Figure 7:
Magnetoresistance as a function of magnetic field at 2, 8, 15 and 35 K for $Gd_3Ru_4Al_{12}$.

**Fig. 1**

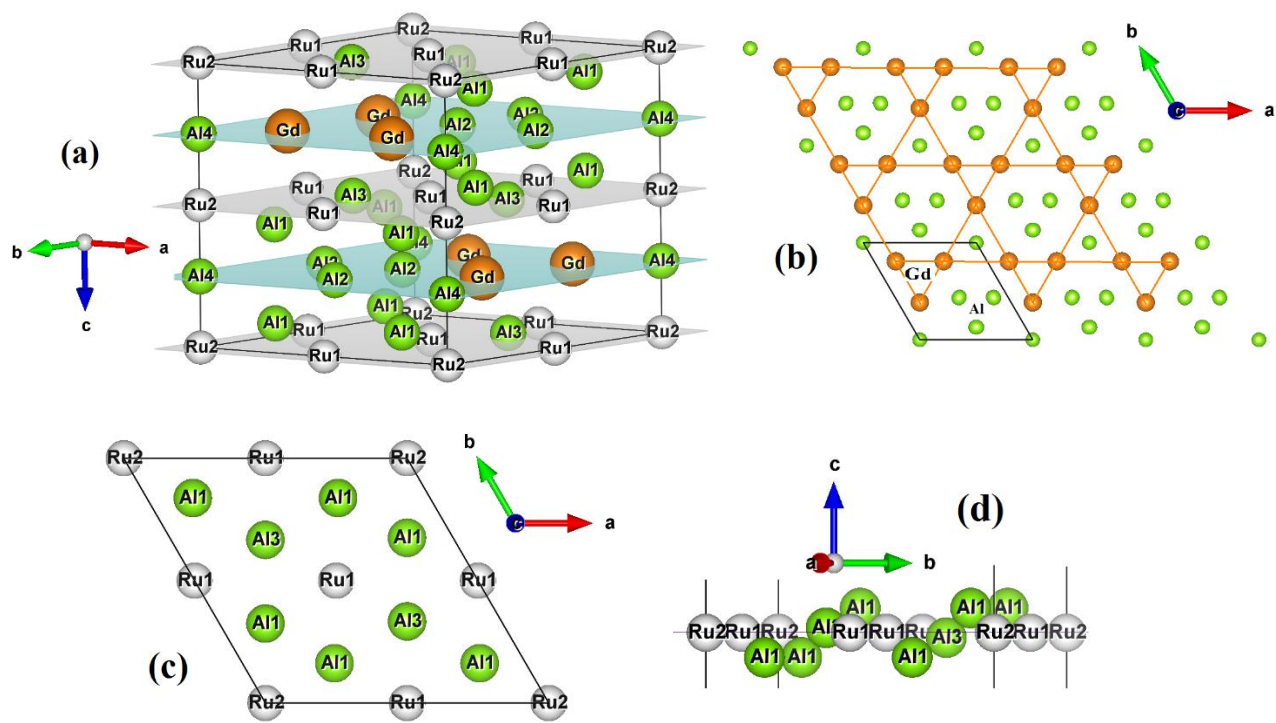

**Fig. 2**

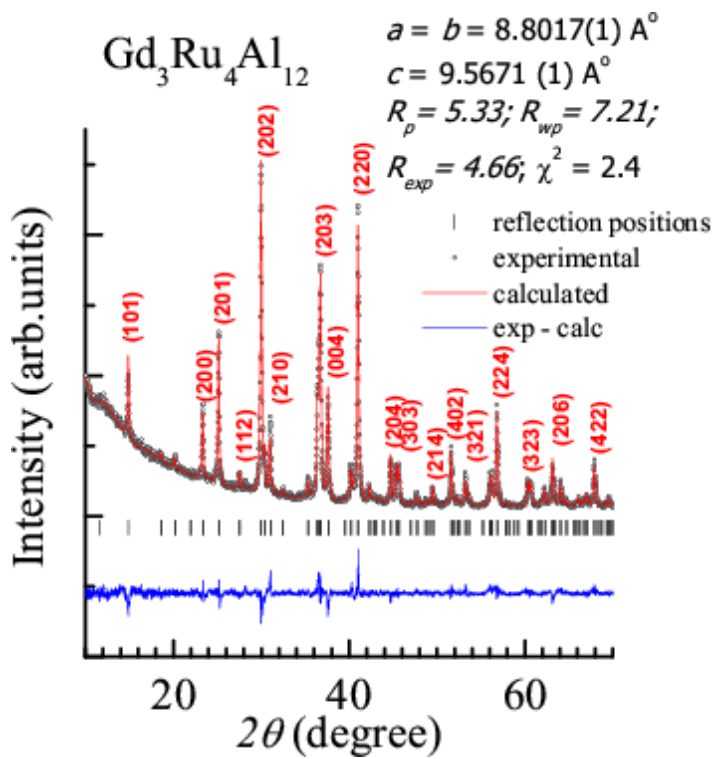

**Fig. 3**

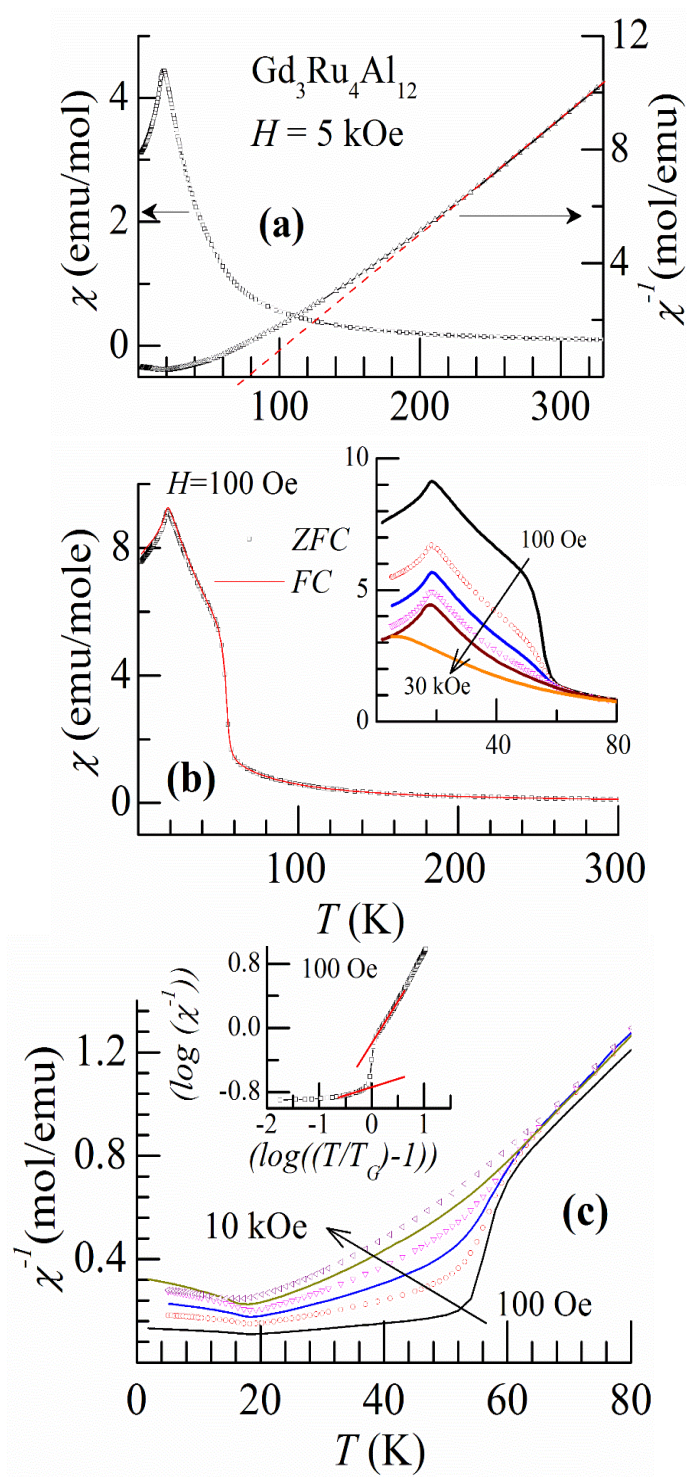

**Fig. 4**

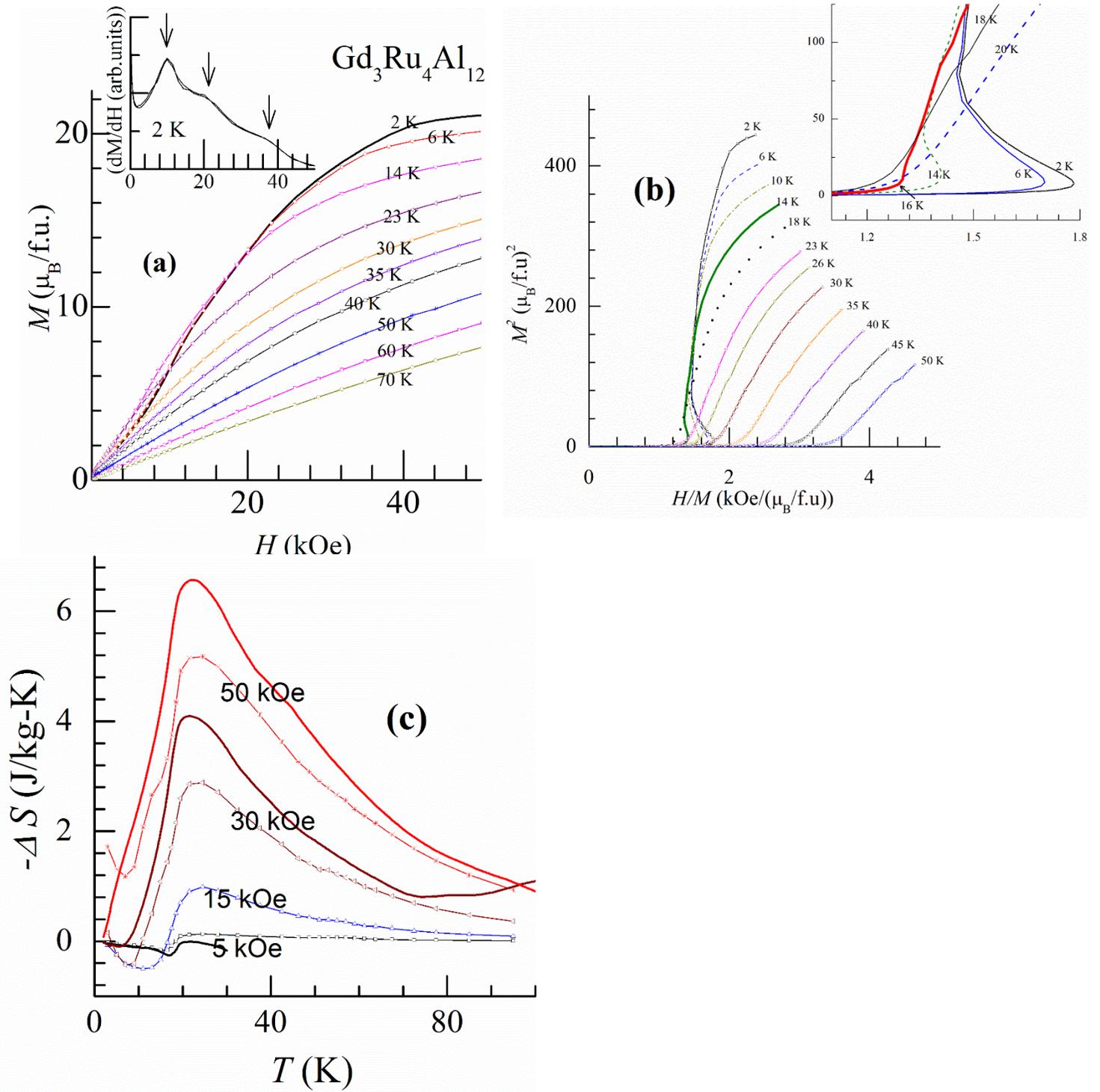

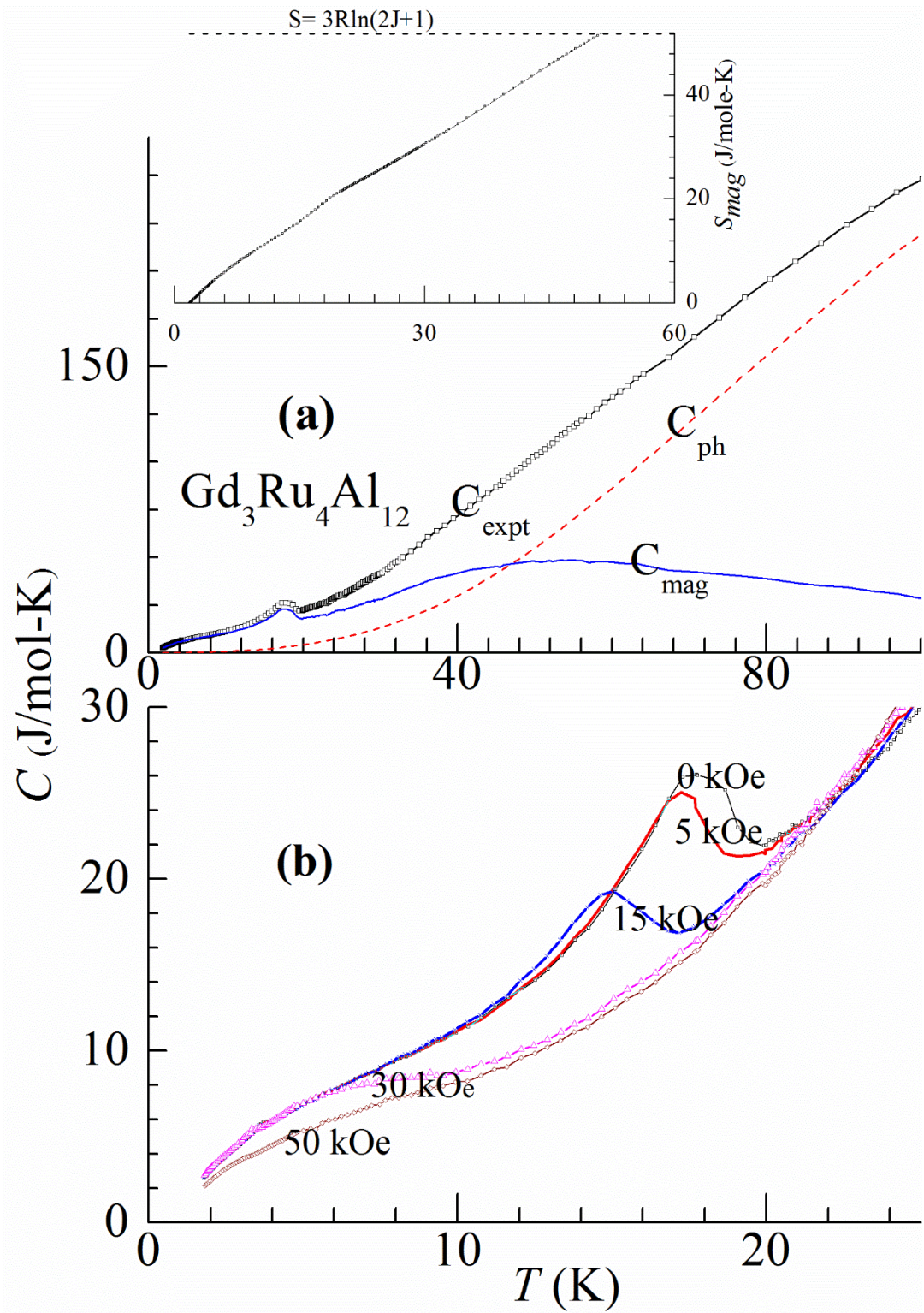

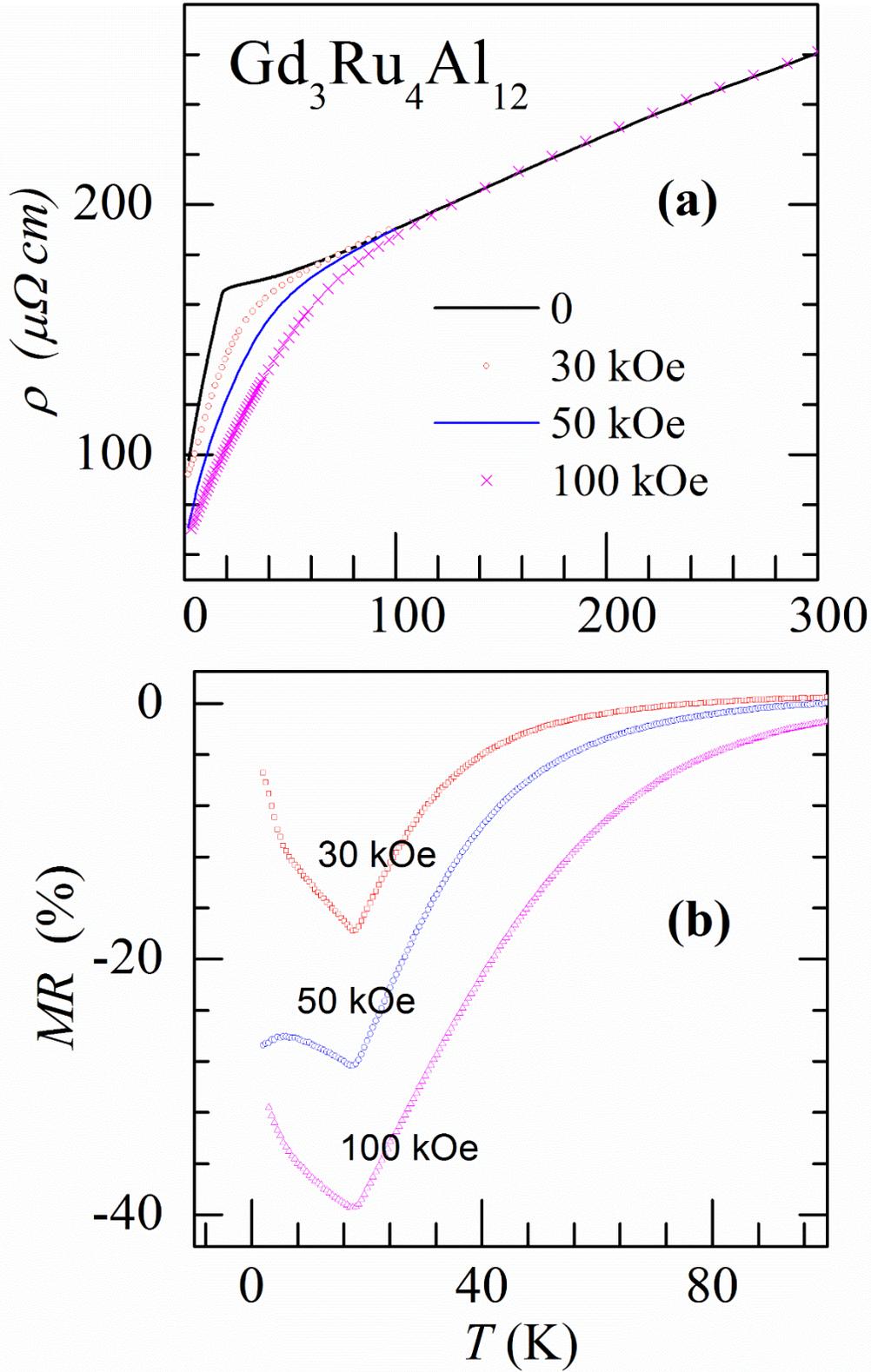

**Fig. 7**

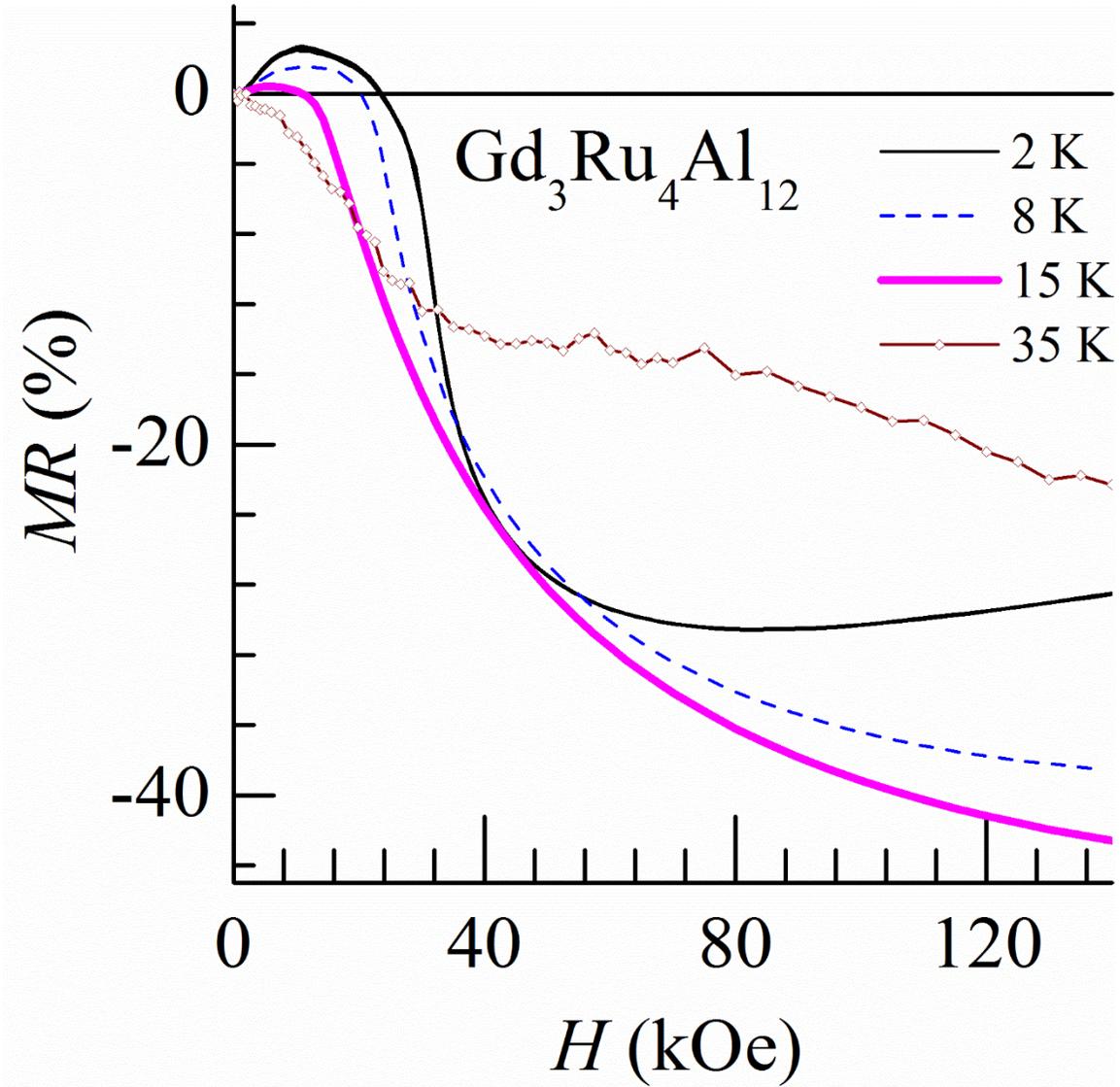